\documentclass[11pt,a4paper]{article}
\usepackage[T1]{fontenc}
\usepackage[utf8]{inputenc}
\usepackage{amsmath,amssymb}
\usepackage{booktabs}
\usepackage{geometry}
\usepackage{xcolor}
\usepackage{listings}
\usepackage{url}
\usepackage[colorlinks=true,linkcolor=blue,citecolor=blue,urlcolor=blue]{hyperref}
\usepackage[round]{natbib}

\newcommand{\pkg}[1]{\textbf{#1}}

\newcommand{\code}[1]{\texttt{#1}}
\newcommand{\proglang}[1]{\textsf{#1}}

\lstdefinestyle{Rinput}{
  language=R,
  basicstyle=\ttfamily\small,
  keywordstyle=\color{blue!70!black},
  commentstyle=\color{gray},
  stringstyle=\color{red!60!black},
  breaklines=true,
  showstringspaces=false,
  frame=single,
  rulecolor=\color{gray!50},
  backgroundcolor=\color{gray!5}
}
\lstdefinestyle{Routput}{
  basicstyle=\ttfamily\small,
  breaklines=true,
  showstringspaces=false,
  frame=single,
  rulecolor=\color{gray!50},
  backgroundcolor=\color{blue!3}
}
\lstnewenvironment{Rinput}{\lstset{style=Rinput}}{}
\lstnewenvironment{Routput}{\lstset{style=Routput}}{}

\title{\pkg{netimpute}: Joint Multiple Imputation of Node Attributes and\\
Network Ties in \proglang{R}}

\author{Robert W.\ Krause}

\date{}

\begin{document}
\maketitle

\begin{abstract}
Missing data in social network studies routinely affects both node-level
attributes and the network ties themselves. Standard multiple imputation software such as
\pkg{mice} \citep{vanBuuren2011} handles the former well but has no notion of
network structure, while dedicated network-imputation procedures typically
treat tie imputation and attribute imputation as separate problems. This
paper introduces \pkg{netimpute}, an \proglang{R} package that (a) computes a
broad battery of node-level structural and attribute
homophily measures across one or more networks, (b) can reduce these to a
manageable set of principal-component predictors while optionally preserving
specific raw measures that are themselves part of substantive hypotheses,
(c) provides a dyadic (cell-level) regression for network ties in the
spirit of MR-QAP \citep{Krackhardt1988}, and (d) combines both into \code{netmice()}, a chained-equations
routine that jointly imputes missing node attributes and missing network ties
by cycling between them, using network-derived predictors for attributes and
attribute-derived (and other-network-derived) predictors for ties. Network
ties are updated \emph{tie-wise} by default: a sequential Gibbs step redraws
each missing tie one at a time, conditional on all previously imputed ties,
refreshing the endogenous statistics (reciprocity, shared contacts) after
every single draw via change statistics. The package additionally supports
structural zeros, logical constraints between networks, social-relations-model
random intercepts in the tie model, custom imputation models with
interactions among any internally created terms, and network-aware per-target
predictor selection (\code{netquickpred()}). We describe the package's design
choices, the measures it computes for binary and non-negative weighted
networks, and illustrate its use with a reproducible example.
\end{abstract}


\section{Introduction}
\label{sec:intro}

Two kinds of missingness arise in studies involving networks, which are
usually
handled by different tools. Missing values on \emph{node attributes}
(age, performance, survey responses, etc.) are the domain of general-purpose
multiple
imputation (MI) software, most prominently \pkg{mice}
\citep{vanBuuren2011}, which implements the fully conditional
specification / chained equations approach to MI
\citep{Rubin1987,LittleRubin2002} or model based likelihood treatments (e.g., FIML). Missing \emph{network ties} are instead the subject of a smaller,
more specialized literature on network imputation
\citep{Huisman2009,Koskinen2010,Krause2018,Krause2020}, often developed
alongside exponential random
graph models \citep{Robins2007} or Stochastic Actor-oriented Models
\citep{Snijders1996,Snijders2017} rather than as an extension of standard MI. A
notable exceptions to this separation are \citet{Koskinen2013} and
\citet{Krause2019}, which both propose joint imputation of attributes and ties,
but neither is implemented as a general-purpose, formula-extensible \proglang{R}
package. They are also limited in the number of network relations and attributes
they can handle.

In practice, these two kinds of missingness are rarely independent
problems. A respondent who did not answer the attribute survey is often the
same respondent whose outgoing nominations are also missing; and the
node-level measures researchers use to \emph{explain} attributes (e.g., is a
central employee also a high performer?) are themselves undefined until the
network is complete, while the covariates used to predict a missing tie
(e.g., do two students share a gender or interest?)
are undefined until the attributes are complete. Imputing one without
regard to the other risks two failure modes: treating an incomplete network
as if it were complete when computing network derived predictors for
attribute imputation, and treating incomplete attributes as if observed when
building the dyadic covariates used to impute ties. Any imputation focused only on one side of this problem works only conditional on the other missing data being irrelevant.

\pkg{netimpute} addresses this by combining the imputation of attributes and network ties together in one function. While the imputation of
the networks is not as thorough as a full model-based approach (e.g., using
SAOMs or ERGMs), it is designed to be fast and scalable to multiple networks and
attributes, while still capturing the essential dependencies between them.

The remainder of this paper is organised as follows. Section~\ref{sec:related}
briefly situates \pkg{netimpute} relative to existing \proglang{R} tools for
network analysis and for missing data. Section~\ref{sec:package} describes
the package's building blocks: node-level measures with a selectable
measure set (\code{net\_measures()} and its \code{net\_measures\_core()} /
\code{net\_measures\_full()} wrappers), predictor construction for
imputation (\code{net\_predictors()}), dyadic regression
(\code{dyad\_regression()}), per-target predictor selection
(\code{netquickpred()}), and the joint imputation routine
(\code{netmice()}). Section~\ref{sec:example} works through a reproducible
example. Section~\ref{sec:discussion} discusses current limitations and
directions for future work.

\section{Related work}
\label{sec:related}

\proglang{R}'s network analysis ecosystem is mature: \pkg{igraph}
\citep{Csardi2006} and \pkg{sna} \citep{Butts2008} both provide extensive
node- and network-level measures, and \pkg{netimpute} builds directly on
\pkg{igraph} for its graph representation and uses \pkg{sna} as an optional
source of a handful of additional centrality measures (Gil-Schmidt power,
flow/load/stress centrality, information centrality, and prestige) that
\pkg{igraph} does not provide. Neither package, however, is designed for
missing data: both assume a fully observed network. The same is true for most functions in the new \pkg{manynet} package \citep{manynet} and its sister packages.

On the imputation side, \pkg{mice} \citep{vanBuuren2011} is the standard
\proglang{R} implementation of chained-equations MI and is used internally
by \pkg{netimpute} as the engine for the actual univariate imputation
step. By default it uses predictive mean matching for numeric and binary targets,
with \pkg{mice}'s other numeric-response univariate methods selectable
through the same \code{method} argument and multinomial
logistic regression via \code{mice::mice.impute.polyreg()}, which in turn
relies on \pkg{nnet} \citep{Venables2002}, for nominal attributes with more
than two categories.

Prior work on missing network data \citep{Huisman2009,Krause2020} has
established that simple treatments (listwise deletion, treating a missing
tie as absent) can materially bias network-level statistics, and has
proposed imputation procedures based on the observed structure. These
procedures are
not implemented as a general-purpose, formula-extensible \proglang{R}
packages; \pkg{netimpute}'s dyadic regression step can be seen as a simple, point-prediction analogue of MR-QAP
\citep{Krackhardt1988} used specifically to build a predictive (rather than
inferential, and thus avoiding the need for permutation) model for tie imputation. The default tie-wise updating scheme
inside \code{netmice()} (Section~\ref{sec:netmice}) additionally borrows
from the exponential-random-graph-model tradition \citep{Robins2007}: each
missing tie is redrawn from its full conditional distribution given the
current state of every other tie --- a Gibbs update of an ERGM-style model
whose parameters are estimated by pseudo-likelihood --- while remaining
restricted to the missing cells, so observed ties are never resampled. This is not seen as a replacement or improvement upon proper ERGM-based (or SAOM-based) imputations, but rather as a fast, scalable, and easy to use way to capture the essential dependencies between ties, attributes, and networks for the purpose of imputation.

\section{The \pkg{netimpute} package}
\label{sec:package}

\subsection{Node-level measures}
\label{sec:measures}

\code{net\_measures\_core(net, attributes, attr\_types = NULL, id\_col =
NULL)} computes nine structural measures per node --- out-degree,
in-degree, a reciprocity measure, Bonacich power centrality (with an
automatically rescaled exponent so it converges regardless of the network's
spectral radius), normalised betweenness, an isolate indicator, Burt's
constraint \citep{Burt1992}, harmonic closeness, and local clustering ---
together with, for every \code{attribute}, a
scale-appropriate homophily/alter-similarity block: an E-I index, Blau's
index, and the modal alter category for binary or multinomial attributes;
average absolute difference on outgoing and on incoming ties, the
mean, minimum, and maximum alter value, plus the mean alter value over
incoming and over outgoing ties separately, for continuous attributes.
\code{net\_measures\_full()} extends this to 28 structural measures
(the full degree family, eigenvector/alpha/PageRank centrality, structural
holes measures, coreness, and, if \pkg{sna} is installed, its six
additional measures) using exactly the same homophily block.

Both functions are wrappers around  \code{net\_measures(net, attributes, measure\_set)}: \code{measure\_set}
accepts any mix of the named sets \code{"core"} and \code{"full"}, the
per-attribute \code{"homophily"} block, and individual measure names, and
returns their union --- e.g.\ \code{measure\_set = c("core",
"total\_degree", "pagerank")} computes the core battery plus two extras,
while \code{c("indegree", "betweenness")} computes exactly those two
measures and nothing else.
Every implemented measure is enumerated in the \code{net\_measures()} and
\code{netmice()} help pages. The same \code{measure\_set} argument is
accepted by \code{net\_predictors()}, \code{netquickpred()}, and
\code{netmice()}, which use these measures as attribute-imputation
predictors.

Only binary (0/1) and non-negative weighted networks are currently
supported. Supplying a network with a negative tie value raises an
error and it is currently recommended to split signed networks into two non-negative networks (e.g., ``friendship'' and
``antagonism'') for separate imputation, and recombine them afterwards
(Section~\ref{sec:discussion}). For non-negative weighted networks, two
measures are redefined:
reciprocity of a node becomes the average $|w_{ij} - w_{ji}|$ over alters tied in
either direction, and local clustering becomes the geometric-mean
weighted clustering coefficient of \citet{OpsahlPanzarasa2009}: for node
$i$ and each pair of its neighbours $(j,k)$, the two-path value
$\sqrt{w_{ij} w_{ik}}$ is summed over every such pair to form a denominator,
and over only the pairs where the triangle is actually closed ($j$ and $k$
are themselves tied) to form the numerator. This ratio reduces exactly to
the  binary clustering coefficient when all weights equal $1$.

\subsection{Predictors for imputation}
\label{sec:predictors}

\code{net\_predictors()} applies \code{net\_measures\_core()} or
\code{net\_measures\_full()} across a list of networks (and a matching list
of attribute data frames), stacks the results, and
offers three output modes: \code{output = "measures"} returns the stacked
raw measures. \code{output = "pca"} instead returns the first
\code{n\_components} principal components of the measure matrix, addressing the multicollinearity that is
otherwise inevitable once a dozen or more structural and homophily measures
are computed per network and the inflation of predictors, which can quickly exceed the $N$ in small networks. \code{output = "both"} returns the principal
components \emph{and} a user-specified set of raw measures
(\code{keep\_vars}), preserved untransformed; this is for the situation
where a specific measure --- say, out-degree --- is itself part of a
hypothesised relationship (i.e., a predictor in a regression) the analyst does
not want diluted by including it
into a composite component, which can also be resiudalised on the remaining predictors so the
components capture variance orthogonal to it (\code{residualize\_kept = TRUE}) .

\subsection{Dyadic regression}
\label{sec:dyad}

\code{dyad\_regression()} vectorises a target network's adjacency matrix
(excluding the diagonal) and regresses it on: for every node attribute, an
ego value, an alter value, and either an absolute difference (continuous
attributes) or a same-category indicator (categorical attributes); reciprocity (the
transposed target network); a bounded $0/1$ two-path indicator
(\code{twopath} --- ``at least one shared contact'', i.e.\ at least one
path $i \to k \to j$); and, for every
\emph{other} network supplied, its tie value, its reciprocity (transpose), and
the sender's and the receiver's out- and in-degree in that network. When
many other networks are supplied, \code{other\_net\_predictors = "pca"}
replaces the node-level degree terms with the first \code{n\_components}
principal components; by default the dyad-level cross-network terms (the other
network's cell and its transpose) always keep their coefficients,
to preserve the direct relationship between the networks. The complete list of term names, as referenced from
\code{netmice()}'s \code{models} formulas, is documented in the
\code{dyad\_regression()} help page. Dyads marked \emph{structurally
absent} (zero by design, e.g., nominations the study design made
impossible; argument \code{structural}) are removed internally from the dyad data
entirely, like the diagonal. The model is fit by \code{glm()} with a user-specified
\code{family} (default Gaussian). Alternatively, \code{random\_intercepts} fits the model
with \pkg{lme4}, adding social-relations-model-style random intercepts per
sender (\code{"ego"}), per receiver (\code{"alter"}), and/or per unordered
pair (\code{"dyad"}; directed networks only --- for an undirected target
the two rows of a dyad are duplicates and the intercept degenerates, which
triggers a warning, and the fixed reciprocity term is dropped when
\code{"dyad"} is used since the dyad effect \emph{is} the SRM reciprocity
term and is exactly singular with it). Unlike in MR-QAP \citep{Krackhardt1988},
permutation is not necessary for this regression, since the goal here is
prediction for imputation
rather than a permutation-based significance test/inference.

\subsection{Per-target predictor selection}
\label{sec:netquickpred}

\code{netquickpred()} is a network-aware, recursive generalisation of
\code{mice::quickpred()}. For every target with missing values it screens
candidate predictors by absolute pairwise-complete correlation --- with the
target's observed values and, like \code{mice::quickpred()}, with its missingness
indicator --- against a threshold \code{mincor} (default $0.1$). For
attribute targets the candidates are the other attributes plus every
node-level network measure of Section~\ref{sec:measures} (computed on the
observed ties); for network targets, screening happens at the dyad level on
the design terms of Section~\ref{sec:dyad}, with the target's own
\code{reciprocity} and \code{twopath} always included unscreened.

Because a
selected predictor's own missing values are imputed inside the chain, the
selection then recurses: with \code{steps = 3} (default), the direct
predictors, their predictors, and the predictors of those all enter the
target's model. Finally, near-duplicate survivors are pruned
per target, either by pairwise correlation (default, threshold $0.9$) or by
variance-inflation factor (threshold $10$), with a network target's
endogenous predictors exempt from pruning. The result
object reports, for every target, exactly which predictors were selected at
which step and with what screening correlation; handed to \code{netmice()}
(or requested there directly via \code{predictor\_selection =
"quickpred"}), it replaces the everything-plus-PCA default of
Section~\ref{sec:netmice} with target-specific models in which every
coefficient keeps its own name. A companion argument, \code{targets}, names
the variables/networks whose imputations are actually needed: everything
else with missing data that is not selected as a (recursive) predictor, target, or explicitly included in an imputation model is
dropped from the imputation algorithm entirely, saving computation time and avoiding that the model gets flooded with irrelevat predictors.

\subsection{Joint imputation}
\label{sec:netmice}

\code{netmice()} is a chained-equations routine, architecturally modelled on
\pkg{mice}'s own sampler, generalised to cycle through both attributes and
networks. Within one iteration, every attribute and every network with
missing values is visited exactly once, in a single interleaved sequence
whose order is randomised independently for each of the $m$ imputation
chains. Visiting an attribute rebuilds its network-derived predictors
(Section~\ref{sec:measures}) from whichever networks were most recently
updated; visiting a network rebuilds its dyadic predictors
(Section~\ref{sec:dyad}) from whichever attributes and other networks were
most recently updated. Because these predictors depend on \emph{the other}
data type, they are never cached: a variable that depends on both a network
and an attribute is effectively recomputed at both points in the cycle.

Each chain starts from a conservative state: missing attribute
values are initialised by resampling observed values, but missing ties are
initialised to $0$ (``no tie''; \code{net\_init = "zero"}), so the first
sweep's endogenous predictors --- reciprocity, two-paths, other networks'
tie and degree terms --- are built from observed ties only. This biases the
very first sweep's structural terms downward, but avoids seeding a
self-reinforcing surplus of random initial ties that later sweeps amplify. A density-based initial imputation is available as \code{net\_init = "sample"}.

\paragraph{Tie-wise updating.} A network visit updates its missing cells
\emph{tie-wise} by default (\code{net\_update = "gibbs"}). A working model
is fit once per visit --- logistic regression for a
binary network, linear regression otherwise --- and one coefficient vector
is drawn from its asymptotic posterior, mirroring \pkg{mice}'s Bayesian
$\beta$-draw so that between-imputation variability reflects estimation
uncertainty. The missing cells are then visited one at a time in random
order; each cell's linear predictor is evaluated with the \emph{current}
values of the endogenous statistics. A binary tie is then drawn from
$\mathrm{Bernoulli}(\mathrm{logit}^{-1}(\eta))$; a weighted tie by
predictive mean matching against the observed dyads' fitted values. Each
draw therefore conditions on all previously imputed ties: this is exactly
the full-conditional (Gibbs) update of an ERGM whose parameters were
estimated by pseudo-likelihood restricted to the missing cells. It yields properly dependent within-network
imputations and avoids potential synchronous-update artifacts of the alternative
scheme, \code{net\_update = "simultaneous"}, in which one model
per visit imputes every missing cell at once from the same snapshot
of the statistics (e.g., if both $x_{ij}$ and $x_{ji}$ are missing, their imputation would be independent of each other).

Under the simultaneous scheme, the tie working model can instead be a
linear mixed model (\code{net\_random\_intercepts}): random intercepts per
sender, receiver, and/or unordered dyad, fit with \pkg{lme4}, capture
actor-level activity/popularity heterogeneity beyond the degree-based fixed
effects, as in the social relations model, with missing ties imputed by
predictive mean matching on the conditional fitted values. Random
intercepts require the simultaneous scheme: setting
\code{net\_random\_intercepts} while leaving \code{net\_update} at its
default falls back to \code{"simultaneous"} with a message, whereas
combining it with an explicit \code{net\_update = "gibbs"} is an error. Note of course that using random intercepts is only meaningful if missing ties are tie non-response and not actor non-response \citep{huisman2008}, since the latter means that all data for the random intercept of a missing node are missing.

\paragraph{Structural zeros and between-network constraints.} Two further
arguments restrict where ties may exist at all. \code{structural} marks
cells that are zero \emph{by design} (e.g., nominations into another school
class that respondents could not make): such cells are excluded from the initialisation donor pool,
and removed from the dyad-level regression rows, so design-zeros cannot
inflate the zeros of the model. \code{net\_dependence} declares
\emph{adaptive} logical constraints between networks: \code{necessary}
rules (``a tie in $A$ is required for a tie in $B$'') and \code{forbidden}
rules (``ties in $A$ and $B$ are mutually exclusive''). Before the chains,
missing cells already determined by observed cells are deduced and filled
(with observed contradictions raising an error); at every visit of a
constrained network, its cells currently determined to be $0$ by the
partners' \emph{current} --- possibly imputed --- state are treated exactly
like structural cells for that visit (i.e., fixed to zero and removed from the estimation); and after every network
re-imputation the rules are followed, so every completed
network list satisfies every rule.

\paragraph{Attributes.} An attribute's predictor set includes the homophily block built from the
target attribute itself, not only from the other attributes: under
homophily or social influence, the composition of a node's alters on the
 variable being imputed (its mean alter value, and in particular the
mean over \emph{incoming} ties, which rest on other --- typically observed
--- nodes' reports) is among the strongest predictors available. The only
measures withheld when imputing an attribute are those that are direct
functions of ego's own current value of that attribute (its E-I index and
the average-absolute-difference terms), since regressing a variable on
a transform of itself would anchor imputed values to their own previous
draws rather than to information from the rest of the data.

The actual univariate draw for an \emph{attribute} is delegated to
\pkg{mice}, generically: \code{method} names any of \pkg{mice}'s
numeric-response univariate methods and is dispatched to the corresponding
\code{mice.impute.<method>()} function --- predictive mean matching
(\code{"pmm"}, the default), its distance-weighted variant
\code{"midastouch"}, random sampling from the observed values
(\code{"sample"}), classification and regression trees (\code{"cart"}),
random forests (\code{"rf"}; requires \pkg{ranger} or \pkg{randomForest}),
the Bayesian and bootstrap linear-regression family (\code{"norm"},
\code{"norm.nob"}, \code{"norm.boot"}, \code{"norm.predict"}), and
\code{"mean"} (the latter two are deterministic and therefore improper for
multiple imputation, as \pkg{mice}'s own documentation cautions).
\code{method} may also be a named character vector assigning
\emph{per-target} methods, \pkg{mice}-style: named entries set the method
for individual attributes or networks, and at most one unnamed entry sets
the default for the rest --- e.g.\ \code{method = c("cart", performance =
"norm")} imputes \code{performance} by Bayesian linear regression and
everything else by trees. Entries naming a multinomial attribute, or a
network whose ties are updated tie-wise or with random intercepts, are
ignored with a message, since those targets' working models are fixed by
design.

Because the auto-generated predictor set can plausibly approach or exceed
the number of observed cases available for a given target --- several
networks' worth of structural and homophily measures adds up quickly
relative to a modest number of nodes --- \code{netmice()} follows the same variable reductions as described above.

A specific, hypothesised relationship can be protected from dilution or
omission via the \code{models} argument, a list of formula strings whose
left-hand side names the attribute or network they apply to; the
right-hand side is evaluated with \code{model.matrix()} (so interaction
syntax such as \code{performance * gender} works) and its columns are
\emph{appended to}, never substituted for, the auto-generated predictors.
A network's own formula must reference the dyadic predictor names from
Section~\ref{sec:dyad} (e.g., \code{age\_absdiff}, \code{advice\_tie}), not
raw node attributes, and may interact any internally created term with any
other --- including the target's own endogenous terms and other networks'
terms: \code{friends \~{} age\_ego:reciprocity}, for instance, lets the
tendency to reciprocate depend on the sender's age. Under the default
tie-wise updating, an interaction involving \code{reciprocity} or
\code{twopath} is, like those main effects, re-evaluated from the current
matrix state after every single tie draw, so the interaction stays live
throughout the sequential sweep. The full lists of referenceable dyad-level
term names and of implemented network measures are given in dedicated
sections of the \code{netmice()} help page. Variables and networks a
formula depends on are automatically protected from being dropped by
the \code{targets} mechanism, so the formula stays evaluable at every
visit.

For $\code{ncores} > 1$, the $m$ chains run as independent \pkg{future}
futures (\code{future::multisession}), each with its own random visit order
and an explicitly pinned random-number generator so that results are
identical whether run sequentially or in parallel given the same
\code{seed} (multisession workers otherwise default to a different RNG
algorithm for proper parallel streams, which would silently change results
if the seed value were reused as-is). \code{netmice()} returns an object of
class \code{netmids} recording, in parallel to \pkg{mice}'s own convergence
diagnostics, the mean and variance of each imputed attribute
value across iterations, and, for every network, its density, reciprocity,
transitivity, isolate count, and average inverse geodesic distance at every
iteration.

\section{Illustrative example}
\label{sec:example}

The example below simulates a 40-node organisation with two networks (a
binary friendship network and a Poisson-weighted advice network) and four
attributes spanning all three supported scales (continuous, binary,
multinomial). Numbers below are the output of running this code with
\proglang{R}~4.6.1 and \pkg{netimpute}~0.1.0; the full script (which also
shows the informational messages about inferred attribute types omitted
here) ships as \code{paper/example\_run.R}. Output lines longer than the
page width are re-wrapped for display.

\begin{Rinput}
set.seed(20260701)
n <- 40
friends <- matrix(rbinom(n * n, 1, 0.08), n, n); diag(friends) <- 0
advice  <- matrix(rpois(n * n, 0.35), n, n);     diag(advice)  <- 0

attrs <- data.frame(
  age         = round(rnorm(n, 34, 7)),
  gender      = sample(c("F", "M"), n, replace = TRUE),
  department  = sample(c("sales", "eng", "hr"), n, replace = TRUE,
                        prob = c(0.4, 0.4, 0.2)),
  performance = round(rnorm(n, 50, 10), 1)
)

core_out <- net_measures_core(friends, attrs)
head(core_out[c("node_id", "outdegree", "indegree",
                "reciprocity_ratio", "betweenness", "constraint")])
\end{Rinput}

\begin{Routput}
  node_id outdegree indegree reciprocity_ratio betweenness constraint
1       1         5        4             0.125       0.095      0.154
2       2         3        1             0.000       0.047      0.302
3       3         2        1             0.000       0.005      0.333
4       4         3        3             0.000       0.062      0.187
5       5         1        0             0.000       0.000      1.000
6       6         2        3             0.000       0.048      0.200
\end{Routput}

Any other subset of the implemented measures can be requested through
\code{net\_measures()} directly (attributes are only needed when the
homophily block is among the requested measures):

\begin{Rinput}
names(net_measures(friends, measure_set = c("indegree", "betweenness",
                                            "pagerank")))
\end{Rinput}

\begin{Routput}
[1] "node_id"     "indegree"    "betweenness" "pagerank"
\end{Routput}

A dyadic regression of friendship ties on the node attributes plus the
advice network illustrates the predictor set from
Section~\ref{sec:dyad} (selected coefficients shown; this network was
simulated with no true structure, so none of these are significant, as
expected):

\begin{Rinput}
dr <- dyad_regression(list(friends = friends, advice = advice),
                       attrs, target = "friends")
summary(dr$model)$coefficients[
  c("age_absdiff", "department_same", "advice_tie"), ]
\end{Rinput}

\begin{Routput}
                     Estimate  Std. Error     t value  Pr(>|t|)
age_absdiff      0.0001873604 0.001025267  0.18274298 0.8550238
department_same -0.0085523890 0.014095721 -0.60673652 0.5441154
advice_tie      -0.0009255629 0.011364711 -0.08144184 0.9351012
\end{Routput}

We now introduce missingness in three attributes and in five percent of the
friendship network's off-diagonal cells, and impute both jointly. The
friendship network's missing ties are updated tie-wise (the default,
Section~\ref{sec:netmice}). \code{models} protects two hypothesised
relationships: an interaction between age and department in predicting
performance and, on the network side, an interaction between the sender's
age and the endogenous reciprocity term --- does the tendency to
reciprocate depend on the sender's age? --- which the tie-wise updater
re-evaluates after every single draw:

\begin{Rinput}
attrs_miss <- attrs
set.seed(1)
attrs_miss$age[sample(n, 4)]         <- NA
attrs_miss$performance[sample(n, 5)] <- NA
attrs_miss$department[sample(n, 3)]  <- NA

friends_miss <- friends
off <- which(row(friends_miss) != col(friends_miss))
friends_miss[sample(off, round(length(off) * 0.05))] <- NA

## maxit = 5 (below the default of 20) keeps the tables below compact.
fit <- netmice(
  attrs_miss, list(friends = friends_miss, advice = advice),
  m = 5, maxit = 5, donors = 5, seed = 20260701, printFlag = FALSE,
  models = list("performance ~ friends_indegree + age * department",
                "friends ~ age_ego:reciprocity")
)
fit
\end{Rinput}

\begin{Routput}
Class: netmids
Imputations (m): 5   Iterations (maxit): 5   Method: pmm   Donors: 5   Cores: 1
Variables with missingness: age, department, performance
Networks with missingness: friends
Network-tie updating: sequential Gibbs (single-tie draws with change
statistics)
Custom models for: performance, friends
\end{Routput}

A completed data set is extracted with \code{complete\_netmice()}, and
convergence can be inspected via the chain means for each imputed attribute
across iterations, alongside network-level diagnostics tracked at every
iteration for every network:

\begin{Rinput}
completed <- complete_netmice(fit, 1)
head(completed$data)
anyNA(completed$data)
anyNA(completed$net_list$friends[row(friends) != col(friends)])
round(fit$chainMean["age", , ], 2)
round(t(fit$netChain["friends", , , 1]), 3)
\end{Rinput}

\begin{Routput}
  age gender department performance
1  31      F      sales        56.3
2  35      M         hr        57.9
3  28      M      sales        57.6
4  13      F         hr        56.1
5  24      M        eng        51.6
6  49      M        eng        42.4
[1] FALSE
[1] FALSE
      1     2     3     4     5
1 33.75 32.25 32.00 36.00 35.25
2 31.75 36.00 32.50 28.50 33.75
3 39.00 36.25 31.00 32.25 33.25
4 34.00 29.00 29.00 32.75 33.00
5 27.50 34.00 31.25 32.25 34.75

  density reciprocity transitivity n_isolates avg_inv_geodesic
1   0.072       0.036        0.126          0            0.504
2   0.070       0.037        0.139          0            0.496
3   0.069       0.056        0.138          0            0.492
4   0.069       0.037        0.137          0            0.495
5   0.072       0.036        0.134          0            0.500
\end{Routput}

Both attributes and network ties are complete in the returned object, and
the tracked network diagnostics are stable across iterations, consistent
with (though of course not a substitute for a formal test of) convergence.

Finally, \code{netquickpred()} (Section~\ref{sec:netquickpred}) shows
which predictors would enter each model under per-target selection. With
only $n = 40$ nodes, chance correlations of network measures easily exceed
the \code{mincor} default of $0.1$, so a deliberately strict threshold
keeps the illustration honest:

\begin{Rinput}
qp <- netquickpred(attrs_miss,
                   list(friends = friends_miss, advice = advice),
                   targets = "performance", mincor = 0.4)
qp
\end{Rinput}

\begin{Routput}
Class: netquickpred (per-target predictor selection)
mincor: 0.4  steps: 3  collinearity: pairwise (threshold 0.9)
missingness screen: TRUE
Targets: performance
Dropped (missing data, not needed): department
- performance (attribute): 0 attribute(s) + 5 network feature(s)
    friends_indegree [step 1], friends_age_alter_min [step 1],
    friends_performance_alter_mean [step 1],
    friends_performance_alter_mean_in [step 1],
    friends_performance_alter_mean_out [step 1]
- age (attribute): 0 attribute(s)
- friends (network): 2 dyad term(s)
    reciprocity, twopath
\end{Routput}

The selection illustrates three mechanisms at once. \code{performance}'s
own alter means --- including the mean over \emph{incoming} ties --- are
selected (the homophily signal discussed in
Section~\ref{sec:netmice}). \code{age} is kept (and will be imputed, here
from an intercept-only model at this strict threshold) because one of
\code{performance}'s selected features, \code{friends\_age\_alter\_min},
derives from it. And \code{department}, whose missing values are needed by
no selected predictor of the target, is dropped from the imputation
entirely. Handing \code{qp} (or simply \code{predictor\_selection =
"quickpred"}) to \code{netmice()} runs the imputation with these lean
per-target models. However, if the \code{netmice()} call above would be repeated
with this \code{qp} object, \code{department} would not be dropped, because it
is part of the predictors in the \code{models} formula for \code{performance}
and is thus protected.

\section{Discussion and future work}
\label{sec:discussion}

\textbf{Signed networks are not yet supported.} Several of the measures in
Section~\ref{sec:measures} are path-based (betweenness, closeness) and
require non-negative edges, while others (the weighted reciprocity and
clustering formulas above) assume weights are magnitudes rather than
signed valences; \pkg{netimpute} currently rejects any network with a
negative tie value. Extending the package to signed networks would
require, at minimum, separate treatment of positive and negative sub-graphs
for the path-based measures and a structural-balance-aware notion of
reciprocity and clustering, which we leave to future work. Until then, it is
recommended that users interested in signed networks split their network into a
positive and a negative side and use the \code{net\_dependence} argument  with
\code{forbidden} for the positive and negative network sides.

\textbf{Endogenous tie statistics.} The tie-wise updater currently
conditions on two endogenous statistics, reciprocity and the bounded
two-path indicator, recomputed from the partly imputed matrices at every
visit. Natural extensions include geometrically down-weighted shared-partner statistics in the spirit
of GWESP terms from the ERGM literature \citep{Robins2007}, and
degree-preserving donor matching for the simultaneous scheme. Overall, the
network imputation is simple and efficient and likely sufficient in many
applications. However, a proper network-model-based imputation (e.g., using ERGM
or SAOM), will be superior, because it is able to incorporate many more network
statistics directly into the imputation model. Obtaining such an imputation at
every iteration of the chained-equations loop is possible but enormously time
consuming and is thus not implemented or suggested here.

\textbf{Method tuning parameters.} \code{netmice()}'s \code{method}
argument dispatches generically to \pkg{mice}'s numeric-response
univariate methods, globally or per target
(Section~\ref{sec:netmice}), but the methods' tuning parameters
(\code{cart}'s \code{minbucket}, \code{rf}'s \code{ntree}, PMM's matching
type) currently keep \pkg{mice}'s defaults; exposing them --- per method,
or per target alongside the method itself --- is a natural next step, as
the internal dispatch is already method-agnostic.

\section{Summary}
\label{sec:summary}

\pkg{netimpute} provides a single, coherent \proglang{R} interface for a
problem that is otherwise split across separate tools, jointly imputing missing
attributes and missing ties through a single chained-equations loop in
which ties are, by default, redrawn one at a time conditional on the
current state of everything else.

\bibliographystyle{apalike}
\bibliography{netimpute}

\end{document}